\input harvmac
\input epsf

\Title{\vbox{\rightline{EFI-97-21}\rightline{hep-th/9704134}}}
{\vbox{\centerline{On the Entropy of Matrix Black Holes}}}
\vskip20pt
\centerline{Miao Li~~ {\it and}~~ Emil Martinec} 
\bigskip
\centerline{\sl Enrico Fermi Inst. and Dept. of Physics}
\centerline{\sl University of Chicago}
\centerline{\sl 5640 S. Ellis Ave., Chicago, IL 60637, USA}

\vskip 2cm



%
%

\def\b0{\bar{0}}
\def\b4{\bar{4}}
%
%
%
\def\journal#1&#2(#3){\unskip, \sl #1\ \bf #2 \rm(19#3) }
\def\andjournal#1&#2(#3){\sl #1~\bf #2 \rm (19#3) }

\def\ie{{\it i.e.}}

\def\cf{{\it c.f.}}

\def\sst{\scriptscriptstyle}

\def\frac#1#2{{#1\over#2}}
\def\coeff#1#2{{\textstyle{#1\over #2}}}

\def\inbar{\,\vrule height1.5ex width.4pt depth0pt}
\def\IC{\relax\hbox{$\inbar\kern-.3em{\rm C}$}}
\def\IR{\relax{\rm I\kern-.18em R}}
\def\IP{\relax{\rm I\kern-.18em P}}

%
%

\def\npb#1#2#3{Nucl. Phys. {\bf B#1} (#2) #3}

\def\plb#1#2#3{Phys. Lett. {\bf #1B} (#2) #3}
\def\prl#1#2#3{Phys. Rev. Lett. {\bf #1} (#2) #3}

\def\prd#1#2#3{Phys. Rev. {\bf D#1} (#2) #3}

\catcode`\@=11
\def\slash#1{\mathord{\mathpalette\c@ncel{#1}}}
\overfullrule=0pt

\def\PP{{\cal P}}

\def\WW{{\cal W}}

\def\lam{\lambda}

\def\underrel#1\over#2{\mathrel{\mathop{\kern\z@#1}\limits_{#2}}}

\catcode`\@=12


%

\def \sinh{{\rm sinh}}
\def \cosh{{\rm cosh}}

\def\sh{{\rm sh}}
\def\ch{{\rm ch}}

\def\bmatrix#1{\left[\matrix{#1}\right]}

\def\aleff{{\alpha'_{\rm eff}}}
\def\lpl{l_p}
\def\gym{g_{\sst\rm YM}}
%
\nref\bfss{T. Banks, W. Fischler, S.H. Shenker and L. Susskind,
hep-th/9610043.}
\nref\bhreview{See for example
J. Preskill, hep-th/9209058, talk at the International
Symposium on Black holes, Membranes, Wormholes and Superstrings, 
Woodlands, TX, 16-18 Jan 1992; 
R. Wald, gr-qc/9702022, talk presented at the Symposium 
on Black Holes and Relativistic Stars 
(dedicated to the memory of S. Chandrasekhar),
Chicago, IL, 14-15 Dec 1996; G. Horowitz, gr-qc/9604051,
talk given at Pacific Conference on Gravitation and Cosmology, 
Seoul, Korea, 1-6 Feb 1996;
J. Maldacena, PhD thesis, hep-th/9607235.
}
\nref\stromvafa{A. Strominger and C. Vafa, hep-th/9601029, 
\plb{379}{1996}{99}.}
\nref\cm{C. Callan and J. Maldacena, hep-th/9602043, \npb{472}{1996}{591}.}
\nref\horstrom{G. Horowitz and A. Strominger, hep-th/9602051; 
\prl{77}{1996}{2368}.}
\nref\dasmathur{S.R. Das and S.D. Mathur, hep-th/9606185,
\npb{478}{1996}{561}.}
\nref\dmw{A. Dhar, G. Mandal, and S. Wadia, hep-th/9605234;
\plb{388}{1996}{51}.}
\nref\maldstrom{J. Maldacena and A. Strominger, hep-th/9609026,
\prd{55}{1997}{861}.}
\nref\limart{M. Li and E. Martinec, hep-th/9703211.}
\nref\dvv{R. Dijkgraaf, E. Verlinde, and H. Verlinde, hep-th/9704018.}
\nref\maldacena{J. Maldacena, hep-th/9605016, \npb{477}{1996}{168}.}
\nref\tseytlin{A.A. Tseytlin, hep-th/9604035, \npb{475}{1996}{149}.}
\nref\cvt{M. Cvetic and A. Tseytlin, hep-th/9606033,
\npb{478}{1996}{181}.}
\nref\duff{M.J. Duff, H. Lu, and C.N. Pope, hep-th/9604052,
\plb{382}{1996}{73}.}
\nref\hms{G. Horowitz, J. Maldacena and A. Strominger, hep-th/9603109;
\plb{383}{1996}{151}.}
\nref\taylor{W. Taylor, hep-th/9611042.}
\nref\larsen{F. Larsen, hep-th/9702153.}
\nref\shrink{W. Fischler, E. Halyo, A. Rajaraman, and L. Susskind,
hep-th/9703102.}
\nref\halyo{E. Halyo, hep-th/9704086.}
\nref\berkrozali{M. Berkooz and M. Rozali, hep-th/9704089.}
\nref\distler{D. Berenstein, R. Corrado, and J. Distler, hep-th/9704087.}
\nref\witten{E. Witten, hep-th/9507121, talk at Strings '95.}
\nref\strominger{A. Strominger, hep-th/9512059, \plb{383}{1996}{44}.}
\nref\ganor{O. Ganor, hep-th/9605201, \npb{489}{1997}{95}.}
\nref\maldsuss{J. Maldacena and L. Susskind, hep-th/9604042,
\npb{475}{1996}{679}.}
\nref\hankleb{A. Hanany and I. Klebanov, hep-th/9606136,
\npb{482}{1996}{105}.}
\nref\klebtseyt{I. Klebanov and A.A. Tseytlin, hep-th/9607107,
\npb{479}{1996}{319}.}
\nref\motl{L. Motl, hep-th/9701025.}
\nref\bankseib{T. Banks and N. Seiberg, hep-th/9702187.}
\nref\dvvstring{R. Dijkgraaf, E. Verlinde, and H. Verlinde, 
hep-th/9703030.}
\nref\gubs{S. Gubser and I.R. Klebanov, hep-th/9609076,
\prl{77}{4491}{1996}.}

%

\newsec{Introduction}

Matrix theory \bfss\ appears to provide a nonperturbative
definition of quantum gravity.  As such, it ought to
resolve the conceptual issues surrounding the quantum mechanics
of black holes \bhreview.  Four- and five-dimensional black
holes with four and three charges $Q_i$, respectively, 
appear to be ideal testbeds in this regard.
Near extremality and in the weak coupling limit $g_{str}Q_i\ll 1$,
it has been shown that configurations of D-branes can account for the 
Bekenstein-Hawking entropy \stromvafa, 
Hawking temperature \refs{\cm,\horstrom}, 
emission and absorption cross-sections \refs{\dasmathur,\dmw}, 
and grey-body factors \maldstrom.  
The five-dimensional case
in particular was examined from the point of view of matrix
theory in \refs{\limart,\dvv}.  In \limart, the authors
analyzed the behavior of probes in the background geometry,
and gathered evidence for an appealing scenario for 
black hole dynamics in matrix theory; closely related
ideas were presented in \dvv, where contact was made with
the weak-coupling D-brane calculations mentioned above.
In this note, we present an alternative derivation of the
Bekenstein-Hawking entropy of these 5d black holes,
using only a plausible assumption about the light degrees
of freedom of $5+1$ super Yang-Mills on a five-torus,
and the special kinematics laid out in \limart.
Borrowing an analysis of this situation by Maldacena
\maldacena\ (which arises in the context of a D-brane
calculation), the entropy follows immediately.
We argue that the leading term in the
entropy is protected by nonrenormalization theorems.

\newsec{The classical solution}

The 5d black hole with three charges can be realized as the
dimensional reduction of a 6d black string, where one of the
charges is momentum travelling down the string (\cf\ \horstrom\
and references therein).
This black string with travelling wave can be realized in
M-theory as a collection of intersecting 2-branes and 5-branes,
with gravitational waves (which we shall call 0-branes)
bound to the 1d intersection \tseytlin.  In an obvious notation,
the configuration is 
\eqn\config{\bmatrix{.&6&7&8&9&11    \cr
           .&.&.&.&.&p_{11}\cr
           5&.&.&.&.&11    \cr }\ .
}
The general nonextremal metric is \refs{\horstrom,\cvt,\duff}
\eqn\fiveM{\eqalign{
  ds^2= T^{1/3}F^{2/3}[T^{-1} & F^{-1}(-K^{-1}h\;dt^2 
		+ K \widehat{dx_{11}}^2)
	+T^{-1}dx_5^2 + F^{-1}(dx_6^2+...+dx_9^2)\cr
	& h^{-1} dr^2 + r^2 d\Omega_3^2]\ ,
}}
where $\widehat{dx_{11}}=dx_{11}+(K'-1)dt$, and
$r^2=x_1^2+...+x_4^2$.  The various functions entering \fiveM\ are
\eqn\various{\eqalign{
  K=&1+\frac{Q_0}{r^2}\qquad,\qquad\qquad Q_0=r_0^2\sinh^2\alpha\cr
  K'=&1-\frac{Q_0^\prime}{r^2}K^{-1} ,\qquad\qquad 
	Q_0^\prime=r_0^2\sinh\alpha\;\cosh\alpha\cr
  T=&1+\frac{Q_2}{r^2}\qquad,\qquad\qquad Q_2=r_0^2\sinh^2\sigma\cr
  F=&1+\frac{Q_5}{r^2}\qquad,\qquad\qquad Q_5=r_0^2\sinh^2\gamma\cr
  h=&1-\frac{r_0^2}{r^2}\ .
}}
The extremal limit is $r_0\rightarrow 0$, $\alpha,\gamma,\sigma
\rightarrow\infty$, with the charges held fixed.
Dimensional reduction gives the 5d Einstein metric
\eqn\fived{
  ds_5^2= -\lam^2(r)h(r)dt^2+\lam^{-1}(r)[h^{-1}(r)dr^2
	+r^2d\Omega_3^2]\ ,
}
where
\eqn\lamdef{
  \lam(r)=(TFK)^{-1/3}=
	\frac{r^2}{[(r^2+Q_0^2)(r^2+Q_2^2)(r^2+Q_5^2)]^{1/3}}\ .
}
Metrics such as \fiveM\ were given a brane interpretation in
\hms, which proposed an identification of the various charges
with numbers of branes and antibranes (translated here into
M-theory equivalents)
\eqn\charges{\eqalign{
  N_0=&\frac{VR_5R^2}{4\lpl^9} r_0^2 e^{2\alpha}\ ,\qquad\qquad
	N_{\bar 0}=\frac{VR_5R^2}{4\lpl^9} r_0^2 e^{-2\alpha}\cr
  N_2=&\frac{V}{4\lpl^6} r_0^2 e^{2\sigma}\qquad,\qquad\qquad
	N_{\bar 2}=\frac{V}{4\lpl^6} r_0^2 e^{-2\sigma}\cr
  N_5=&\frac{R_5}{4\lpl^3} r_0^2 e^{2\gamma}\qquad,\qquad\qquad
	N_{\bar 5}=\frac{R_5}{4\lpl^3} r_0^2 e^{-2\gamma}\ .\cr
}}
Here $V$ is the volume of the four-torus spanning $x_6,...,x_9$;
$R_5$ is the radius of the $x_5$ circle; and $R$ is
the radius of the $x_{11}$ circle.  A useful relation is
\eqn\rzero{
  r_0^2=4R^2\left(\frac{N_2N_{\bar 2}\;N_5N_{\bar 5}}{N_0N_{\bar 0}}
	\right)^{1/2}\ ,
}
allowing one to relate the different brane numbers in 
an interesting way:
\eqn\balance{
  \left(\frac{N_0\lpl}{R}\right)
	\left(\frac{N_{\bar 0}\lpl}{R}\right) =
  \left(\frac{N_2 RR_5}{\lpl^2}\right)
	\left(\frac{N_{\bar 2} RR_5}{\lpl^2}\right) =
  \left(\frac{N_5 RV}{\lpl^5}\right)
	\left(\frac{N_{\bar5} RV}{\lpl^5}\right)\ .
}
The first of these quantities has an obvious interpretation --
it is the invariant mass (in Planck units) of the gas of
massless gravitons on the intersection string.  Since the other
two are U-dual to a gas of massless particles, we will call them
the `invariant masses' of the two-branes and five-branes.
Equation \balance\ then says that there is an equipartition
of `invariant masses' or component energies of the black hole.
In \refs{\limart,\dvv}, it was argued that the proper way to embed these
solutions in matrix theory is to take the light-front
directions $x^\pm=t\pm x_{11}$, and boost to the infinite
momentum frame (IMF).  In this process, $N\equiv N_0\rightarrow\infty$ 
and $N_{\bar 0}\rightarrow0$. 

\newsec{Matrix realization and state counting}

Matrix theory compactified on a transverse $T^5$ is thought to be
equivalent to $5+1$ super Yang-Mills theory on the dual torus
${\tilde T}^5$ \refs{\bfss,\taylor}.  In this construction, a five-brane
wrapped on the longitudinal direction is an instanton;
a longitudinal membrane is represented by momentum flux
$T_{0i}$; and $N$ units of 0-brane charge are realized
using $U(N)$ as the Yang-Mills gauge group.
For the configuration \config, we take the instanton flux along
$6789$, and the momentum flux in the 5 direction.
The instanton in the $5+1$ gauge theory is a solitonic string, whose
collective coordinates consist of 4 bosons and 4 fermions
describing the transverse oscillations.  The instanton
charge on a torus can split into $N$ pieces, hence 
the effective string tension is decreased by a factor of $N$.
For sufficiently large values of the charges, these `instanton strings'
dominate the entropy \maldacena\foot{It seems that the counting
of string states is a common thread in entropy computations;
see \larsen\ for an intriguing discussion and further references.}.  
Because there are both
winding and anti-winding strings (see figure 1a),
representing fivebranes and anti-fivebranes, 
it is possible for an instanton anti-instanton
pair to annihilate.  This takes place locally by a 
joining-splitting interaction, figure 1b.  The winding energy
is temporarily converted into kinetic energy; further 
joinings/splittings will transfer some of this energy back into winding.
The equilibrium configuration will have a string gas
with a rough equipartition between kinetic and potential energy.

\bigskip
{\vbox{{\epsfxsize=4in
        \nobreak
    \centerline{\epsfbox{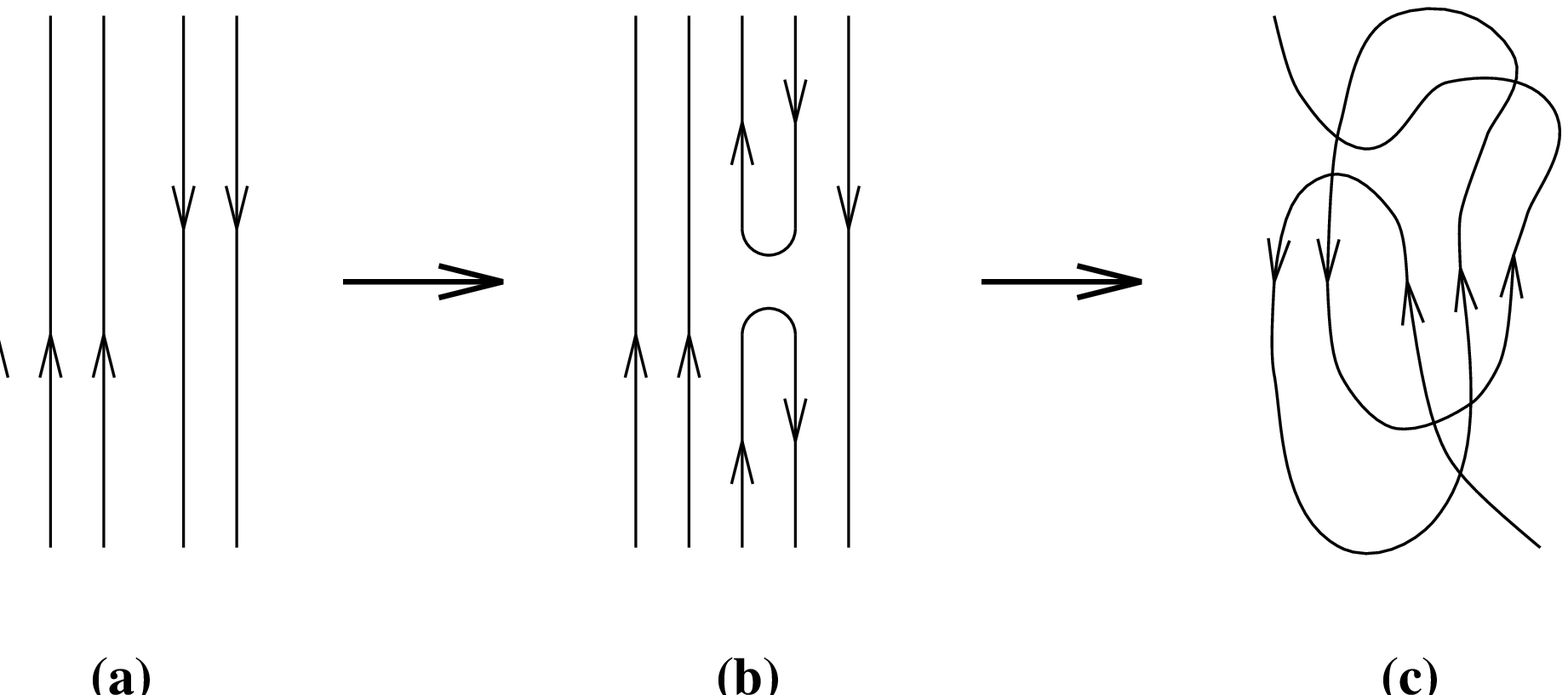}}
        \nobreak\bigskip
    {\raggedright\it \vbox{
{\bf Figure 1.}
{\it Maldacena's picture of the gas of `instanton strings'.
Through repeated joining/splitting interactions, the energy
is collected into the entropically preferred state --
one large string.}
 }}}}
    \bigskip}

The statistically preferred set of configurations (figure 1c)
has a single long string carrying essentially all the energy -- 
the instanton strings
are in the Hagedorn phase (note that the formulation is
microcanonical, so this is well-defined).  
Thus to compute the entropy, we need to determine the number of states
available to a single long string with $c_{\rm eff}=6$,
winding charge along the 5 direction $(N_5-N_{\bar 5})$,
and momentum charge along the 5 direction $(N_2-N_{\bar 2})$.

The ADM energies carried by the various charges are 
\eqn\energies{\eqalign{
  \lpl E_0=&\frac{VRR_5}{4\lpl^8} r_0^2 e^{2\alpha}
	=(\frac{N_0N_{\bar0}\lpl^2}{R^2})^{1/2} e^{2\alpha}\cr
  \lpl E_2=&\frac{VRR_5}{4\lpl^8} r_0^2 e^{2\sigma}
	=(\frac{N_0N_{\bar0}\lpl^2}{R^2})^{1/2} e^{2\sigma}\cr
  \lpl E_5=&\frac{VRR_5}{4\lpl^8} r_0^2 e^{2\gamma}
	=(\frac{N_0N_{\bar0}\lpl^2}{R^2})^{1/2} e^{2\gamma}\ ,\cr
}}
with corresponding antibrane expressions obtained by flipping
the sign in the exponentials.  
The total ADM energy is thus
\eqn\adm{
  \lpl E_{ADM}=(N_0+N_{\bar 0})\frac{\lpl}{R}
	+(N_2+N_{\bar 2})\frac{RR_5}{\lpl^2}
	+(N_5+N_{\bar 5})\frac{RV}{\lpl^5}\ .
}
The energy of the system not
carried by zero-branes is available to the string, 
since the IMF energy equals
\eqn\imfen{
  E_{LC}=p_+= E_{ADM}-\frac{N}{R}\ .
}
Note also that this energy is the Hamiltonian
of the 5+1 gauge theory.  The energy available to oscillators
of the instanton string is reduced by the constraint
that the black hole carry net two-brane and five-brane
charge, which are carried on the string as momentum
$\lpl \PP=(N_2-N_{\bar2})\frac{RR_5}{\lpl^2}$ and winding
$\lpl \WW=(N_5-N_{\bar5})\frac{RV}{\lpl^5}$.
We will defer until the end of this section all questions
about string interactions, as well as the range of parameter
space over which our discussion is valid.
Meanwhile, we will treat the instanton string as noninteracting.
Then the left and right excitation numbers are 
\eqn\nlr{\eqalign{
  n_{L,R}=~&{\aleff}[E_{LC}^2 - (\PP\pm\WW)^2]\cr
	=~&\frac{\aleff}{\lpl^2}\left[\frac{VRR_5}{4\lpl^8}
		\; r_0^2\right]^2
		[(\ch2\sigma+\ch2\gamma)^2-(\sh2\sigma\pm\sh2\gamma)^2]\cr
	=~&\frac{\aleff}{\lpl^2}\left[\frac{VRR_5}{4\lpl^8}
		\; r_0^2 \right]^2 4\ch^2(\sigma\mp\gamma)\ .\cr
}}
The entropy is now evaluated as
\eqn\bhent{\eqalign{
  S_{BH}=~&2\pi\left[\sqrt{\coeff16{c_{\rm eff}}n_L} 
		+ \sqrt{\coeff16{c_{\rm eff}}n_R}\right]\cr
	=~&2\pi\left[\frac{\aleff}{\lpl^2}\cdot
		\frac{VR_5R^2}{\lpl^7}\right]^{1/2}
		(\sqrt{N_2}+\sqrt{N_{\bar2}})
		(\sqrt{N_5}+\sqrt{N_{\bar5}})\ .
}}
One must have $\aleff=\frac{N\lpl^9}{VR_5R^2}$ to match the entropy.
Naturally, the energy per unit length of an instanton string
in $5+1$ gauge theory is 
\eqn\tension{
  T_{\rm eff}=\frac{4\pi^2}{\gym^2 N}=\frac{VR_5R^2}{2\pi N\lpl^9}
}
(the $1/N$ arises from the charge fractionalization mentioned above).
Then with the standard relation $T_{\rm eff}=(2\pi\aleff)^{-1}$,
the Hagedorn gas of instanton strings precisely accounts
for the Bekenstein-Hawking entropy.  
The combination $\gym^2 N$ appearing in \tension\
suggests that conventional large-N techniques might be
useful for the study of the instanton string gas.
It is important to note that the factors 
in the entropy cannot be ascribed to
particular branes/antibranes; everything gets mixed up
in the `plasma' of light excitations, as we see from figure 1.
Another important feature is that the tension $\tension$
is finite in the limit $N,R\rightarrow\infty$,
$N/R^2$ fixed that characterizes the large $N$ limit
with fixed longitudinal momentum density and fixed
entropy per unit length.

It would be interesting to calculate 
the temperature $\beta_H^{-1}$ of Hawking radiation
from the present viewpoint; one has
\eqn\hawktemp{\eqalign{
  \beta_H=~&\frac{\pi}{a}\sqrt{N}(\sqrt{N_2}+\sqrt{N_{\bar2}})
	(\sqrt{N_5}+\sqrt{N_{\bar 5}})  \cr
  \lpl a=~&2 \frac{RR_5}{\lpl^2} \left(N_2N_{\bar2}\right)^{1/2}
	=2\frac{RV}{\lpl^5}\left(N_5N_{\bar5}\right)^{1/2}\ .
}}
Note that the `invariant mass' $2a=T_H S_{BH}$
would appear to be the difference between the energy
and the free energy.
One can compare this temperature to that of the one-dimensional
gas of excitations on the string:
\eqn\Toned{
  \beta_{1d} = 
	\frac{\pi}{2a}\left[
	\frac{\ch\gamma\;\ch\sigma}{\ch 2\gamma+\ch 2\sigma}\right] \ .
}
Using the relation 
$\beta_H=2\pi(\aleff)^{1/2}\;\ch\gamma\;\ch\sigma$,
which may be deduced from \hawktemp, one finds
\eqn\relation{
  \beta_H=\beta_{1d}[4 (\aleff)^{\frac12} E_{LC}]\ .
}
This reduces \maldacena\ to the relation 
$\beta_H=\beta_{\sst\rm Hagedorn}$ in the near-extremal limit.

Another property of the black hole that we can motivate from
the instanton string picture is the relation \balance\
between the invariant masses of two-branes and five-branes.
If in equation \nlr\ we introduce adjustable parameters
for the amount momentum and winding, $\PP=a\;\sinh2\sigma$,
$\WW=b\;\sinh2\gamma$, $E_{LC}=(a\;\cosh2\sigma+b\;\cosh2\gamma)$,
then one has
\eqn\lre{
  E_{L,R}=\left[E_{LC}^2-(\PP\pm \WW)^2\right]
	=\bigl(a^2+b^2+2ab\;\cosh[2(\sigma\mp \gamma)]\bigr)\ .
}
The entropy \bhent\ turns out to depend only
on the ratio $t=a/b$; variation of this parameter gives
\eqn\varia{
  \delta S=\pi\Bigl({1\over \sqrt{n_L}}+{1\over \sqrt{n_R}}\Bigr)
	NN_2N_5e^{-2\sigma-2\gamma}(1-1/t^2)\delta t\ ,
}
which determines the stationary point $t=1$. This is just the
balancing formula
\eqn\pbala{
  RR_5l_p^{-2}\sqrt{N_2N_{\bar{2}}}=RVl_p^{-5}\sqrt{N_5N_{\bar{5}}}\ .
}
However, there is a puzzle: The second derivative $S''(t=1)>0$,
suggesting the system is unstable.  We are confused about the
interpretation of this fact.

Let us pause now to compare the derivation of $S_{BH}$ given
here with that in \dvv.
In \dvv, the $x_5$ circle is shrunk to a sub-Planckian size
in order to recover a matrix description of a regime where
one can compare to string perturbation theory.  The 1+1d
field theory on the dual circle gives a matrix description
of IIA strings bound to NS five-branes.  The infrared description
of this field theory is an N=(4,4) supersymmetric sigma model
on a target space
\eqn\target{
  S^{N_0N_5}T^4\ ,
}
and the entropy is effectively the density of states of this sigma
model at level $N_2$, 
$S_{BH}=\sqrt{N_0N_5}(\sqrt{N_2}+\sqrt{N_{\bar2}})$.
The interactions of the matrix strings are marginal
operators in this effective field theory, but this will not
affect the central charge and hence the leading term in the
entropy.  One is performing a rather similar
calculation as the one above which resulted in equation \bhent.
However, the 1+1 field theory in \dvv\ is restricted to a kind of
static gauge description of the strings, where the spatial
direction along the string is identified with $x_5$.
Our approach is somewhat more covariant in this respect.
This is necessary, since in the Hagedorn regime describing
`fat' black holes (those where all the brane numbers 
\charges\ are macroscopic),
the string wanders ergodically, with many
overhangs so that static gauge breaks down.
One sees from \balance\ that if one takes $R_5$ small and $V$ large, 
then one can have large numbers of (anti)two-branes at minimal
cost; however (anti)five-branes are expensive,
therefore they will annihilate as far as possible.  
In other words, we have recovered the picture of \dvv,
where there are only the instanton strings winding
in one direction (the large dual circle to $x_5$ with
radius $\sim R_5^{-1}$), with a gas of momentum modes
travelling in both directions.  A second possibility
(realizing part of the U-duality of the entropy)
involves taking vanishing size of the transverse $T^4$ of
volume $V$ on which the five-branes are compactified,
while making $R_5$ large.
Then the situation is reversed: The (anti)five-branes
have low cost, while the (anti)two-branes annihilate
as far as possible.  We imagine that other limits might be possible
by exploiting more of the duality group, \cf\
\shrink-\distler.

The central charge and hence the leading term in the entropy
is extremely robust.  We are dealing with the lightest
excitations of a non-BPS state in a maximally
supersymmetric gauge theory.  These light excitations are
described by an action with N=(4,4) 2d supersymmetry, and so the
entropy is protected by nonrenormalization theorems for the
central charge of this theory (in fact the central charge
is quantized for this much supersymmetry).  The effective
string tension, which governs the relation between spacetime
and worldsheet energies, is determined by the $F^2$ term in
the gauge theory and is also not renormalized.
This feature opens up exciting new possibilities for solving
problems such as the cosmological constant.
The vacuum we live in is a nonsupersymmetric state in a
supersymmetric theory; if the effective matrix dynamics is
governed by a supersymmetric theory, the cosmological
constant could be protected even though in most other respects
there is no supersymmetry.

One possible cause for concern in the analysis leading to \bhent\
is the effect of string interactions, which we have been
ignoring.  However, their chief influence is to change the
connectivity of the Hagedorn string locally, without altering
its position (\ie\ the string occupies the same one-dimensional
locus immediately before and after interaction).  Thus,
if anything, the string merely explores its phase space
more rapidly and efficiently; the phase space distribution
and hence the entropy ought not to be significantly altered.
Indeed, in the calculation of \dvv\ it was proposed that the
main effect was simply to shift the target space background
\target\ by resolving its orbifold singularities, leaving
the entropy unchanged.

Until now, we have not discussed the range of validity of
our calculation, for instance how far from extremality
it can be trusted.  
One limitation might come from the UV behavior
of the 5+1 Yang-Mills theory, which is not well understood
(for recent discussions in the context of matrix theory,
see \shrink-\distler).
In the present situation, although the energy density is large,
this is achieved by macroscopically populating the longest
wavelength modes of the system (rather than putting the 
system at finite temperature, for instance, whereby high momentum
modes would be excited).
Therefore we needn't worry about short-distance effects.
One needs to work in a regime where 
there are enough charges $N_2$, $N_{\bar2}$,
$N_5$, $N_{\bar5}$ so that one is in the asymptotic region
of the level density of the instanton string.
From \balance, one sees that $\gym^2N$ becomes large unless
either $R_5$ or $V$ is small, corresponding to the perturbative
regime accessible to D-brane technology.
Thus the difficulty with `fat' black holes 
is not that the Yang-Mills theory is dominated by ultraviolet
behavior, but rather that it becomes strongly coupled in a
different way: $\gym\rightarrow0$, $N\rightarrow\infty$,
but $\gym^2N\rightarrow\infty$.  In this limit, the
tension of the instanton string goes to zero.

The remarkable fact about the entropy formula
\bhent\ is that it seems to be universally valid,
suggesting that the density of states is always 
controlled by a string theory.
At high density, any piece of instanton string finds itself
close to many other pieces.  Now recall that the instanton
string is the matrix theory representation of the wrapped
five-brane; wrapped fivebranes approaching one another
generates `tensionless' strings \refs{\witten,\strominger}
(see \ganor\ for the large $N$ limit).  
Could it be that `tensionless'
here refers merely to the effective tension of a string in its
Hagedorn phase?  In any event, we would like to interpret
the black hole entropy \bhent\ as telling us
that there is a string theory controlling the properties
of large $N$ 5+1 super Yang-Mills theory 
in the regime appropriate to large black holes,
and that \bhent\ is its nonperturbative density of states.
A rigorous demonstration of this fact would show us how string theory
accomodates large black holes as quantum states.
Remarkably, questions about the nature of these objects
have been transformed in the infinite momentum frame into
issues of the renormalization group in field theory.

\newsec{Discussion}

The entropy counting confirms the picture of \limart,
wherein the black hole was viewed as a plasma of excitations
in the super Yang-Mills theory.  The special property
of the 5d black holes considered here is that one can identify
the light excitations of the 5+1 super Yang-Mills plasma
with the collective modes of a string.
It should be possible to count the entropies of other M-theory
black holes (for a survey, see \cvt), provided
we can identify the light excitations of the associated Yang-Mills
theory in states carrying the relevant charges.
In order to embed such black holes in matrix theory, one
wants one of the charges carried by the hole to be momentum,
so that this can be identified with 0-brane charge.

This was seen to be possible for the 4d black hole with
three charges in \limart; black holes in $6\le d\le 9$
can be realized as boosted two-branes \cvt, and hence
also can be realized in matrix theory.
Thus black holes in dimension six and above
might be understood in a similar fashion, by identifying
the relevant soft modes of the Yang-Mills plasma.
The difference with the 4d and 5d cases is that, in $6\le d\le 9$,
the entropy vanishes as one approaches extremality, 
as a power of $r_0$.  
From formulae analogous to \rzero\ and \balance, the entropy will have
to have a power of anti-brane charge in it.
Looking at the ADM mass \cvt
$$E_{ADM}\propto r_0^{D-3}[\coeff{2}{D-3}+
\cosh 2\sigma+\cosh 2\alpha]\ ,$$
it would appear that there is a (fraction of a) constituent for which an
appropriate charge cannot be turned on, or perhaps the constituents
have a nontrivial interaction energy.
It would, of course, be interesting to understand
the specific form of the soft modes of the plasma that gives the
required scaling properties of entropy as a function of mass.

The 4d case is a collection of intersecting five-branes
carrying momentum on their intersection strings.  In the 6+1
Yang-Mills theory on the dual torus, the instantons
describing the longitudinal fivebranes are membrane-like
objects extended over the orthogonal dual coordinates.
It would be interesting to see if one can extract the
entropy from the dynamics of a single space-filling (Hagedorn-like)
membrane carrying the wrapping charges along three mutually 
orthogonal planes.  
It has been suggested by Hanany and Klebanov \hankleb\ 
(see also \klebtseyt)
that the properties of intersecting fivebranes are related
to a noncritical string living in the three-dimensional
intersection.  Here one of these three dimensions is $x_{11}$,
so in the IMF one is looking for a string inside a two-brane.
Perhaps this is the effective object which
controls the density of states.  Note also that it is not
properly understood how to compactify M-theory in this
situation \shrink.  Turning this around, we might use the properties
of black holes to gain a window into the relevant
high-dimensional field theories.

It is important that
the excitations of the system are of long wavelength, and have an
energy scale which is suppressed by $1/N$ \refs{\maldsuss,\shrink},  
\motl-\dvvstring. 
This property is responsible for recovering perturbative
string theory and its exponential density of states 
in the large $N$ limit, when a circle is shrunk to sub-Planckian
size.  It is the collective
motions of the various objects (instantons, torons, etc.)
that dominates the entropy of black holes.  
In order to account for the entropy of fat black holes,
the UV (high frequency) properties
of the system ought never to become important,
or else the system could dissolve into the
dynamics of field theory.
It would be a disaster if that were to happen, as one would
not be able to acount for the exponential density of states
of fat black holes.

One can now see the outline of a calculation of scattering
from a large black hole, once we understand enough about the
5+1 supersymmetric field theory.
The matrices describing the black hole/probe system are
initially on the Coulomb branch, having the approximate structure
\eqn\probe{
\bmatrix{{\bf BH} & 0\cr
	 0 & {\rm probe}}\ .
}
The energy of the probe is in the kinetic energy of
the Higgs field, the lower right block of the matrix.
As the clump of matter comes in, the difference in the 
scalar vevs goes to zero; the energy is absorbed by the
instanton string gas, being deposited in gauge field
modes over the whole matrix.  In the tails of the
probability distribution, it may happen that a fluctuation has
the off-diagonal fields almost vanishing in some row(s)
and column(s), while the diagonal entries of the
scalars have some kinetic energy; a block of the matrix can
then re-emerge onto the Coulumb branch -- in other words,
a Hawking particle escapes. 

There may be some useful analogies to exploit between the
present kinematical setup (black string stretched along the
longitudinal $x^+$--$x^-$ plane) and the parton picture
of QCD.  We have seen that the black hole entropy is
carried by the `instanton string' representing the effects
of 2-brane and 5-brane constituents.  These are like the
valence quarks that carry the quantum numbers of the system
(they carry the entropy and all gauge charges, including
the angular momentum \maldacena).
The 0-branes play the role of `wee partons' \bfss.
The Hawking radiation spectrum consists of such wee partons,
since it is dominated by the emission of massless particles
carrying small longitudinal momentum.
Indeed, calculations of the emission of charged scalars 
\refs{\gubs,\maldstrom}
(here charge is momentum along $x_{11}$) confirm
that the radiation is peaked near zero longitudinal
momentum.  Thus the Hawking radiation
spectrum can be regarded as a prediction of the distribution
of wee partons in the black hole, as a function of longitudinal
momentum fraction.  It would be very interesting to verify
this distribution function by a calculation in the present
context.


\vskip 1cm
\noindent{\bf Acknowledgments:} 
We are extremely grateful to  
T. Banks 
and 
J. Maldacena
for discussions.
EM thanks the Rutgers University theory group for its
hospitality during the course of this work.

This work was supported by DOE grant DE-FG02-90ER-40560 and NSF grant
PHY 91-23780.

\listrefs
\end